# Minimizing Hidden-Node Network Interference by Optimizing SISO and MIMO Spectral Efficiency

D. W. Bliss, *IEEE Senior Member*, S. Govindasamy, *IEEE Member*

### Abstract

In this paper, the optimal spectral efficiency (data rate divided by the message bandwidth) that minimizes the probability of causing disruptive interference for ad hoc wireless networks or cognitive radios is investigated. Two basic problem constraints are considered: a given message size, or fixed data rate. Implicitly, the trade being optimized is between longer transmit duration and wider bandwidth versus higher transmit power. Both single-input single-output (SISO) and multiple-input multiple-output (MIMO) links are considered. Here, a link optimizes its spectral efficiency to be a "good neighbor." The probability of interference is characterized by the probability that the signal power received by a hidden node in a wireless network exceeds some threshold. The optimized spectral efficiency is a function of the transmitter-to-hidden-node channel exponent, exclusively. It is shown that for typical channel exponents a spectral efficiency of slightly greater than 1 b/s/Hz per antenna is optimal. It is also shown that the optimal spectral efficiency is valid in the environment with multiple hidden nodes. Also explicit evaluations of the probability of collisions is presented as a function of spectral efficiency.

## I. INTRODUCTION

The hidden-node problem is a significant issue for ad hoc wireless networks and cognitive radios. This paper extends the discussion on this topic that was presented in [1]. In that paper, the problem of a single fixed message size was considered assuming a capacity achieving link. Here, that analysis is extended, and it is shown that the fixed message size solution applies to the fixed data rate problem given a sufficiently low average data rate. Also, a more realistic approximation to the achievable link spectral efficiency is introduced.

Dan Bliss is currently with MIT Lincoln Laboratory, Lexington, Massachusetts 02420. Siddhartan Govindasamy is with Franklin W. Olin College of Engineering, Needham, MA 02492.

### A. Problem Definition

In general, the problem is characterized by two links and four nodes: a transmitter of interest, a receiver of interest, a hidden transmitter, and hidden receiver [2]. It is assumed that the existence of the hidden link is not known by the link of interest and that the hidden link cannot adapt to interference. The hidden receiver (denoted the hidden node) may be near enough to the transmitter of interest that the interference causes the hidden link to fail. Being a "good neighbor," the link of interest would like to minimize the adverse effects of transmitting a message or frame of some finite number of bits. Both single-input single-output (SISO) and multiple-input multiple-output (MIMO) wireless communication links are considered.

As an example, imagine other links in an ad hoc wireless network (or legacy network for the cognitive radio problem) are communicating with random occupancy in frequency and time, as seen in Figure 1(a). The geometry of a particular transmitter, receiver, and hidden node is depicted in Figure 1(b). The distance from the transmitter to the hidden node is $r$, and the radius $r_i$ contains the region within which the interference-to-noise ratio (INR), denoted $\eta$, is sufficient to disrupt the hidden link.

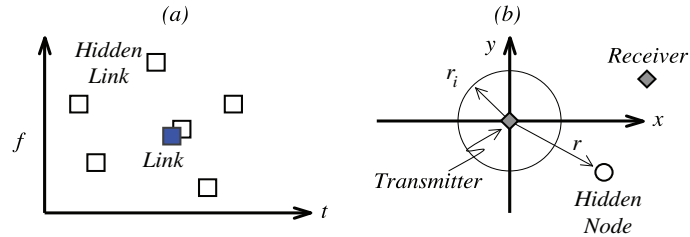

Fig. 1.    (a) Displays a set of notional hidden links in time ($t$) and frequency ($f$) in the presence of a signal-of-interest transmission. (b) Depicts a notional geometry of transmitter, receiver, and hidden node. The distance from the transmitter to the hidden node is given by $r$. The region of disruptive interference is contained with radius $r_i$.

There are six main assumptions used in this analysis. First, the probability of interfering is relatively small, which is equivalent to saying that the spatial, temporal, spectral occupancy of the network is not particularly high near the link of interest so that the probability of multiple collisions can be ignored. Second, the effects of interference on the hidden node can be factored into the probability of collision and the probability that the INR ($\eta$) at the hidden node exceeds some critical threshold $\eta_i$. Third, the hidden node location is sampled uniformly over some large area. Fourth, the average channel attenuation from the transmitter to the hidden node can be accurately modeled by using a power-law attenuation model. Fifth, the link performance can be characterized with reasonable accuracy by the channel capacity as a function of signal-to-noise ratio (SNR) modified by an implementation loss. Sixth, the hidden node does not have some interference mitigation capability that prefers a particular waveform structure.

For the transmitter of interest to cause disruptive interference at the hidden node, the interfering signal must satisfy two requirements. First, it must overlap with a hidden link spectrally and temporally such that transmissions collide. The probability of collision (sufficient overlap) is denoted $p_c$. Second, the interfering signal must be of sufficient strength at the hidden node to cause disruptive interference, assuming sufficient overlap in time and frequency. The probability that the distance from the transmitter to hidden node is within sufficient range to cause disruptive interference is denoted $p_r$. The probability of disruptive interference is denoted $p_i$ and is given by the product of $p_c$ and $p_r$.

### B. Fixed Message Size Versus Average Data Rate

For a link that has a single message of some finite number of bits, the transmitter is assumed to transmit until the completion of the message. It is also assumed that the message occupies some contiguous portion of the spectrum. For a fixed number of bits, the optimization minimizes the probability of interference by varying the time-bandwidth product. Because of the formulation of the problem, the temporal and spectral dimensions can be treated equally.

Under the average data rate constraint, a related problem is considered. For some spectral allocation of bandwidth $B$, a fixed average data rate is requested. If the requested average data rate is very high, then there is little to optimize. The best strategy is to minimize the spectral efficiency by setting the spectral efficiency to the average data rate divided by the bandwidth of the spectral allocation.

At a very low average data rate, the optimal strategy is to transmit with some low occupancy that approximates the fixed message optimization. In this paper, only the low duty cycle regime is considered. As an example consider a periodic temporal solution, so that repetition period $T_0$ (or equivalently in spectral occupancy) is large compared to the lengths of the signal of interest or other links transmission durations. As a more specific example, consider an ad hoc network in which all links randomly and asynchronously communicate within a fixed frequency allocation, and observe the same waveform optimization as seen in Figure 2.

## II. Optimal SISO Spectral Efficiency

For some $n_{\text{info}}$ information bits, the probability of interference $p_i$ is a function of period of transmission $T$ and bandwidth of transmission $B$, $p_i = p_i(T, B, n_{\text{info}})$. This functional dependence is developed by noting that the probability of collisions $p_c$ is dependent upon the transmitted duration and bandwidth.

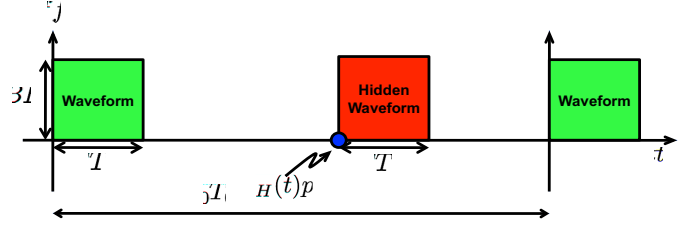

Fig. 2. Jointly optimized ad hoc network with randomly located hidden nodes. To achieve some average data rate, nodes transmit periodically with separation $T_0$ and duration $T$. The probability of the hidden node starting to receive a transmission at time $t$ is given by $p_H(t)$.

## A. Fixed Message Length

For a fixed message length, given the assumption stated in the introduction, this functional dependence is developed by noting that the probability of collisions $p_c$ is linearly related to the transmitted duration and bandwidth

$$p_c \propto (T + T_H)(B + B_H),\tag{1}$$

where $T_H$ and $B_H$ are the temporal and spectral extents of the hidden node link, because the fraction of the temporal and spectral space occupied by the links [seen in Figure 1(a)]. In the case of a cognitive radio that is attempting to transmit a larger message, the probability of collision is approximated well by

$$p_c \propto T\,B\,.\tag{2}$$

and thus the probability of collision is proportional to the area subtended by the link in the temporal-spectral space, under the assumption that the distribution of packet occupancy over frequency and time is uniform. Similarly, in the case of an ad hoc network for which all waveforms are being optimized jointly, the durations of the optimized waveform of interest and the hidden waveforms are the same so that the probability of collision is given by

$$p_c \propto 4\,T\,B \propto T\,B\,.\tag{3}$$

If it is assumed that the hidden node is randomly located with respect to the transmitter in a two-dimensional physical space [Figure 1(b)], then the probability that the hidden node is within sufficient range, $p_r$, to cause disruptive interference is proportional to the area, $A$, over which the signal has a sufficient INR, $\eta > \eta_i$, at the hidden node

$$p_r \propto A(\eta > \eta_i)\,.\tag{4}$$

Consequently, to a good approximation, the probability of interference exceeding some threshold is given by

$$p_i \propto T\,B\,A\,.$$

(5)

The area is a function of the transmit energy and propagation loss to the hidden node.

For a SISO system, the information theoretic bound [3] on the number of bits that can be transmitted within time $T$ and bandwidth $B$ is given by

$$n_{\text{info}} \leq T\,B\,c$$

(6)

$$\tilde{c} = \log_2(1 + \gamma)\,,$$

$$c = \log_2\left(1 + \frac{\gamma}{l}\right)\,,$$

where $\tilde{c}$ is the information theoretic limit in bits/s/Hz on the SISO spectral efficiency (assuming a complex modulation), and $\gamma$ is the signal-to-noise ratio (SNR) at the receiver. The bound is not achievable for finite $n_{\text{info}}$, but it is a reasonable approximation to the limiting performance. To approximate a more realistic rate $c$, it is assumed that the achieved spectral efficiency is given by the information theoretic capacity with an additional implementation loss figure $l$, so that $\gamma \to \gamma/l$.

By assuming the link of interest can be approximated by modified capacity, the SNR at the receiver can be expressed in terms of the number of bits transmitted and the spectral efficiency

$$n_{\text{info}} \approx T\,B\,\log_2\left(1 + \frac{\gamma}{l}\right)\,,$$

$$\gamma \approx l\,(2^c - 1)\,.$$

(7)

If the channel gain to the hidden node is denoted $b^2$ and the channel gain to the receiver of interest is denoted $a^2$, then the INR, denoted $\eta$, at the hidden node is

$$\eta = \frac{b^2}{a^2}\gamma\,.$$

(8)

By using a simple power-law model for loss, with the channel gain to the hidden node proportional to $r^{-\alpha}$, the radius $r_i$ at the critical interference level (at which $\eta = \eta_i$) is found by observing

$$\gamma = \frac{a^2}{b^2}\eta \propto \frac{a^2}{r^{-\alpha}}\eta \Rightarrow \frac{a^2}{r_i^{-\alpha}}\eta_i$$

$$r_i \propto \gamma^{1/\alpha} = l^{1/\alpha}\,(2^c - 1)^{1/\alpha}\,.$$

(9)

Consequently, the probability of interference for the SISO system is given by

$$p_i \propto T \, B \, A$$

$$\approx \frac{n_{\text{info}}}{c} A \propto \frac{n_{\text{info}}}{c} r_i^2$$

$$\propto \frac{(2^c - 1)^{2/\alpha}}{c} \, . \tag{10}$$

The optimal spectral efficiency for some $\alpha$ is given by

$$\frac{\partial p_i}{\partial c} \propto \frac{2^{c+1} \left(-1 + 2^c\right)^{\frac{2}{\alpha}-1} \log(2)}{c\alpha} - \frac{(-1 + 2^c)^{2/\alpha}}{c^2} = 0$$

$$c_{opt} = \frac{\alpha + 2\, W_0\!\left(-\frac{1}{2} e^{-\alpha/2} \alpha\right)}{2 \log(2)} \tag{11}$$

$$\approx 1.355(\alpha - 2) - 0.118(\alpha - 2)^2 + 0.008(\alpha - 2)^3 \, , \tag{12}$$

where $\log(\cdot)$ indicates the natural logarithm, and $W_0(x)$ is the product log or principal value of the Lambert W-function[1] [4]. It is remarkable that optimal spectral efficiency is dependent upon the channel exponent exclusively. It is also interesting to note that a similar result was found when attempting to optimize the spectral partitioning of an interference limited network [5].

In Figure 3, the optimal spectral efficiency for a given channel exponent, under the assumption of ideal coding in a static channel, is displayed. In the absence of multipath scattering, the line-of-sight exponent is $\alpha = 2$ (an anechoic chamber for example). For $\alpha = 2$, the optimal spectral efficiency approaches zero. For most scattering environments, $\alpha = 3$ to $4$ [6] is a more reasonable characterization, suggesting an optimal spectral efficiency around 2 b/s/Hz.

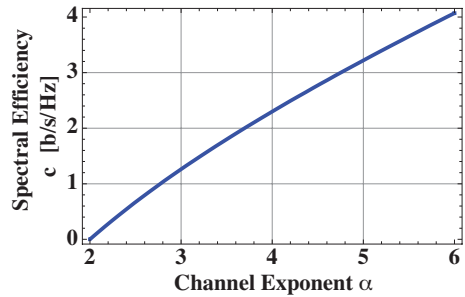

Fig. 3. Optimal SISO spectral efficiency, $c$, for ideal coding in a static environment as a function of transmitter-to-hidden-node channel gain exponent, $\alpha$.

[1] The Lambert W-function is the inverse function of $f(W) = W e^W$. The solution of this function is multiply valued.

## B. Fixed Average Data Rate

For a fixed average data rate given the assumptions stated in the introduction, the solution that minimizes the probability of interference is similar to that of the fixed message length situation if the requested average data rate is not too large, and the transmitter has the ability to change the transmit power and data rate over the duration of a frame. The best solution is then to transmit at the optimal fixed message spectral efficiency during some portion of the frame and to be off during the remaining portion of the frame.

This optimization approach works until the stationary point solution is no longer realizable. As a specific example, consider the spectral allocation filling ad hoc network indicated in Figure 2. In this case, the hidden link durations $T$ are the same as the link of interest. The duty cycle $d$ is then given by

$$d = \frac{T}{T_0} \, . \tag{13}$$

When the spectral efficiency optimization drives the transmit duration $T$ to the point that the sum of the transmit duration and the hidden-link duration fills the repetition period $T + T_H = 2T = T_0$, the optimization has hit a limit and the derivative of the probability and thus the optimization is no longer valid. In this case, the optimization is valid if the duty cycle is less than $d < 1/2$. Consequently, the requested average rate $R$, must be less than

$$R = d \, c_{opt} \, B$$
$$< \frac{1}{2} \, c_{opt} \, B \, , \tag{14}$$

under the spectral filling, equal transmission filling $T = T_H$ assumptions.

## III. POISSON FIELD OF HIDDEN NODES

The analysis of the previous section can be readily extended to networks with multiple hidden nodes and fixed message lengths. Suppose that the hidden nodes are distributed on the plane according to homogenous a Poisson point process (PPP) with average density $\rho$ nodes per unit area. Assume that each node receives packets from a transmitter with an average rate of $\lambda$ packets per unit time and that all packet arrivals and node positions are independent. Additionally, assume that the packets destined for the hidden nodes are negligible in duration compared to the packet duration of the link of interest $T$, and that the rate of packet arrivals to each hidden node is small enough that at most one packet arrives at each hidden node in a time interval $T$.

Since we are only concerned with a time window of duration $T$, we can assume that each hidden node receives at most one packet. This enables us to model the system as a three-dimensional PPP with two dimensions corresponding to the location of the hidden nodes on the plane and the third being the time of packet arrival. Hence, the probability of no collision equals the probability that there is no point of the 3-dimensional PPP in a cylinder of cross-section $A$ and height $T$ whose base is centered at the transmitter of interest as shown in Figure 4. The probability of no points in this cylinder equals $e^{-\lambda \rho A T}$. Hence, the probability of interference is simply

$$p_i = 1 - e^{-\lambda \rho A T}$$

To minimize this probability, $A T$ has to be minimized. This yields the same optimal spectral efficiency $c_{\text{opt}}$ as in the case of a single hidden node. The corresponding probability of interference assuming a data rate equal to the Shannon capacity (i.e. $l = 1$) is simply

$$p_i = 1 - \exp\left( -\lambda \rho \pi r_{link}^2 \frac{n_{\text{info}}}{c_{\text{opt}}} \left( \frac{\sigma^2}{\eta_i} \left( 2^{c_{\text{opt}}} - 1 \right) \right)^{\frac{2}{\alpha}} \right)$$

where $\sigma^2$ is the variance of the noise, $r_{link}$ is the length of the link of interest and $c_{\text{opt}}$ is given by (11). Note that for the PPP model of hidden node locations $\pi \rho r_{link}^2$ equals the average number of hidden nodes closer to the transmitter of interest than its target receiver.

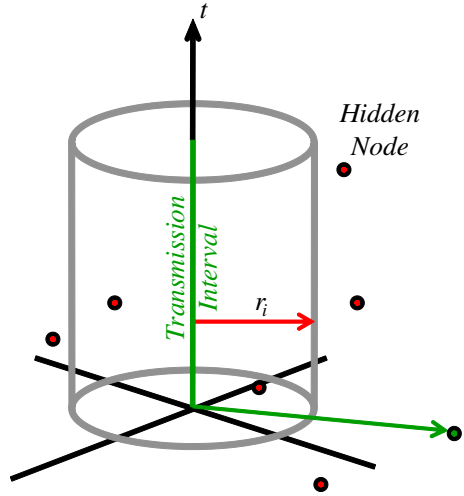

Fig. 4.   Illustration of exclusion region for a Poisson field of hidden nodes.

To illustrate the benefit of spectral-efficiency optimization in Poisson networks, we plotted the probability of interference versus spectral efficiency for such a network with nominal parameters $n_{\text{info}} = 1024$, $r_{link} = 10$, $\rho = 0.001$, $\lambda = 0.001$, $\eta_i = -30$ dB and $\sigma^2 = 10^{-14}$ in Figure 5. Observe that there is an optimal spectral efficiency to minimize probability of collision as predicted and that a significant increase in probability of collision occurs of spectral efficiencies that are too high or low are utilized.

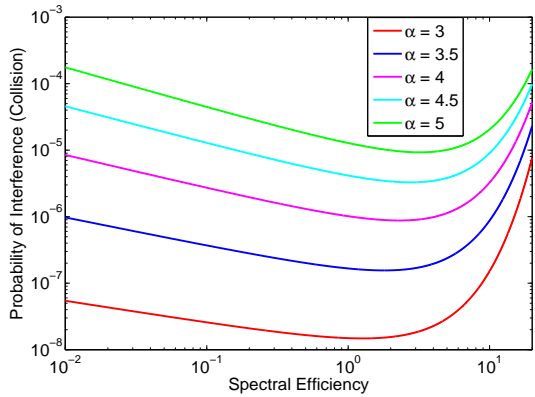

Fig. 5. Probability of collision vs. spectral efficiency for a Poisson field of hidden nodes. The nominal parameters used for this plot are $n_{\text{info}} = 1024$, $r_{link} = 10$, $\rho = 0.001$, $\lambda = 0.001$, $\eta_i = -30$ dB and $\sigma^2 = 10^{-14}$.

## IV. Optimal MIMO Spectral Efficiency

The analysis for a MIMO link is similar to the SISO link.

### A. Fixed Message Length

It is assumed that the MIMO link has an uninformed transmitter (without channel state information), and the number of transmitters by number of receivers, $n_t \times n_r$, MIMO channel is not frequency selective. The received signal is given by $\mathbf{Z} = \mathbf{H}\,\mathbf{S} + \mathbf{N}$, where $\mathbf{Z} \in \mathbb{C}^{n_r \times n_s}$ is the received signal, $\mathbf{S} \in \mathbb{C}^{n_t \times n_s}$ is the transmitted signal, $\mathbf{H} \in \mathbb{C}^{n_r \times n_t}$ is the channel matrix, and $\mathbf{N} \in \mathbb{C}^{n_r \times n_s}$ is the noise. The number of transmitted symbols is $n_s$.

For a MIMO system with an uninformed transmitter (a transmitter without channel state information), the information theoretic bound on the number of bits transmitted is given by

$$n_{\text{info}} \leq T\,B\,c$$

$$c = \log_2 \left| \mathbf{I} + \frac{P_0}{l\,n_t}\,\mathbf{H}\,\mathbf{H}^{\dagger} \right| \tag{15}$$

where $c$ is an approximation of the achievable rate for the MIMO spectral efficiency (assuming a complex modulation), $P_0$ is the total thermal-noise-normalized transmit power, and $l$ is implementation loss figure.

This rate is constructed by modifying the information theoretic capacity with the implementation loss figure. The notation $|\cdot|$ indicates the determinant. Implicit in this formulation is the assumption that the interference-plus-noise covariance matrix is proportional to the identity matrix which is a reasonable model for most interference avoiding protocols.

Because the capacity is a function of a random SNR matrix, there is not a single solution as there is in the SISO analysis. However, by assuming that the channel matrix is proportional to a matrix sampled from an i.i.d. zero-mean element-unit-norm-variance complex Gaussian matrix, $\mathbf{H} = a\mathbf{G}$, an asymptotic analysis, in the limit of a large number of antennas, a solution can be found. Surprisingly, the asymptotic model is a reasonable approximation for even small numbers of antennas [7]. With this model, the term $a^2 P_0$ is the the average SNR per receive antenna at the receiver of interest. To simplify the analysis, it is assumed that $n_r = n_t \equiv n$. The optimal spectral efficiency under the assumption of other ratios of number transmitters to receivers can be found following a similar analysis. The asymptotic capacity [7] is given by

$$
\begin{aligned}
\frac{c}{n} &\approx \frac{a^2 P_0}{l \log 2} {}_3F_2([1,1,3/2],[2,3],-4\,a^2 P_0/l) \ \equiv f(a^2 P_0/l) \\
&= \frac{4 \log\left(\sqrt{4a^2 P_0/l+1}+1\right)}{\log(4)} + \frac{\sqrt{4a^2 P_0/l+1}}{a^2 P_0/l \,\log(4)} \\
&\quad - \frac{1}{a^2 P_0/l\,\log(4)} - 2 - \frac{2}{\log(4)}\,,
\end{aligned}
\tag{16}
$$

where ${}_pF_q$ is the generalized hypergeometric function [8], and the function $f(x)$ is used for notational convenience.

The SNR at the receiver can be expressed in terms of the number of bits transmitted and the spectral efficiency

$$
\begin{aligned}
n_{\text{info}} &= T\,B\,n\,f\left(\frac{a^2 P_0}{l}\right)\,, \\
a^2 P_0 &= l\,f^{-1}\!\left(\frac{c}{n}\right)\,.
\end{aligned}
\tag{17}
$$

Unfortunately, a simple formulation of the functional inverse of $f(x)$ [denoted $f^{-1}(y)$] is not available; however, it is tractable numerically. By using a model similar to the link of interest, if the channel to the hidden node is given by $\mathbf{H}_{hn} = b^2 \mathbf{G}_{hn}$, then the average INR per receive antenna at the hidden node is given by

$$
\eta = b^2 P_0\,.
\tag{18}
$$

By using a similar analysis to the SISO case and power-law model for the average channel gain $b$, the radius of disruptive interference is found by observing

$$a^2 P_0 = \frac{a^2}{b^2} \eta \propto \frac{a^2}{r^{-\alpha}} \eta \Rightarrow \frac{a^2}{r_i^{-\alpha}} \eta_i$$

$$r_i \propto (a^2 P_0)^{1/\alpha} = l^{1/\alpha} \left[ f^{-1}\!\left(\frac{c}{n}\right) \right]^{1/\alpha}. \tag{19}$$

Consequently, the probability of interference for the MIMO system is given by

$$p_i \propto T\,B\,A = \frac{n_{\text{info}}}{c}\,A \propto \frac{n_{\text{info}}}{c}\,r_i^2$$

$$\propto \frac{\left[ f^{-1}(\beta) \right]^{2/\alpha}}{\beta}, \tag{20}$$

where the number-of-antenna-normalized spectral efficiency is given by $\beta \equiv c/n$. The optimal spectral efficiency for a given $\alpha$ is given by

$$\beta_{opt} = \mathrm{argmin}_\beta \frac{\left[ f^{-1}(\beta) \right]^{2/\alpha}}{\beta}$$

$$\approx 0.795(\alpha - 2) + 0.028(\alpha - 2)^2 - 0.003(\alpha - 2)^3. \tag{21}$$

In Figure 6, the optimal spectral efficiency per antenna for a given channel exponent, under the assumption of ideal coding in a static channel, is displayed. For most scattering environments, $\alpha = 3$ to $4$ [6] is a reasonable characterization, suggesting a spectral efficiency per antenna of a little more than $1$ b/s/Hz.

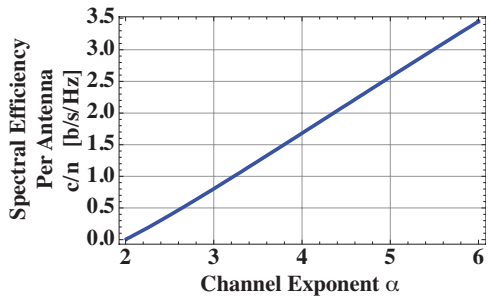

Fig. 6. Optimal MIMO spectral efficiency for ideal coding in a static environment as a function of the transmitter-to-hidden-node channel gain exponent.

## B. Fixed Average Data Rate

The discussion for the fixed average data rate for the MIMO link is essentially the same as that for the SISO link. Under the spectral filling assumption, the duty cycle $d$ and the requested average data rate $R$ of the link are related by

$$R = d\, n\, \beta_{opt}\, B\,. \tag{22}$$

## V. Conclusion

The optimal spectral efficiency was calculated to minimize the probability of causing disruptive interference in a wireless network. The optimization trades the time-bandwidth product versus the power of a transmission. Remarkably, the optimal spectral efficiency is a function of the channel exponent, exclusively. By using this information, the optimal per antenna spectral efficiency falls between 0.5 and 2 for typical channel exponents.

The authors would like to thank Paul Fiore, Peter Parker and Dorothy Ryan for their comments.